\documentclass{aastex631}

\usepackage{xcolor}
\usepackage{multirow}
\usepackage{tabularx}
\usepackage{graphicx}
\usepackage{amsmath}

\begin{document}
\title{Investigating Pulsar Wind Nebula DA~495: Insights from LHAASO and Multi-Wavelength Observations}

\author{Zhen Cao}
\affiliation{State Key Laboratory of Particle Astrophysics \& Experimental Physics Division \& Computing Center, Institute of High Energy Physics, Chinese Academy of Sciences, 100049 Beijing, China}
\affiliation{University of Chinese Academy of Sciences, 100049 Beijing, China}
\affiliation{TIANFU Cosmic Ray Research Center, 610000 Chengdu, Sichuan,  China}
 
\author{F. Aharonian}
\affiliation{TIANFU Cosmic Ray Research Center, 610000 Chengdu, Sichuan,  China}
\affiliation{University of Science and Technology of China, 230026 Hefei, Anhui, China}
\affiliation{Yerevan State University, 1 Alek Manukyan Street, Yerevan 0025, Armeni a}
\affiliation{Max-Planck-Institut for Nuclear Physics, P.O. Box 103980, 69029  Heidelberg, Germany}
 
\author{Y.X. Bai}
\affiliation{State Key Laboratory of Particle Astrophysics \& Experimental Physics Division \& Computing Center, Institute of High Energy Physics, Chinese Academy of Sciences, 100049 Beijing, China}
\affiliation{TIANFU Cosmic Ray Research Center, 610000 Chengdu, Sichuan,  China}
 
\author{Y.W. Bao}
\affiliation{Tsung-Dao Lee Institute \& School of Physics and Astronomy, Shanghai Jiao Tong University, 200240 Shanghai, China}
 
\author{D. Bastieri}
\affiliation{Center for Astrophysics, Guangzhou University, 510006 Guangzhou, Guangdong, China}
 
\author{X.J. Bi}
\affiliation{State Key Laboratory of Particle Astrophysics \& Experimental Physics Division \& Computing Center, Institute of High Energy Physics, Chinese Academy of Sciences, 100049 Beijing, China}
\affiliation{University of Chinese Academy of Sciences, 100049 Beijing, China}
\affiliation{TIANFU Cosmic Ray Research Center, 610000 Chengdu, Sichuan,  China}
 
\author{Y.J. Bi}
\affiliation{State Key Laboratory of Particle Astrophysics \& Experimental Physics Division \& Computing Center, Institute of High Energy Physics, Chinese Academy of Sciences, 100049 Beijing, China}
\affiliation{TIANFU Cosmic Ray Research Center, 610000 Chengdu, Sichuan,  China}
 
\author{W. Bian}
\affiliation{Tsung-Dao Lee Institute \& School of Physics and Astronomy, Shanghai Jiao Tong University, 200240 Shanghai, China}
 
\author{J. Blunier}
\affiliation{APC, Universit\'e Paris Cit\'e, CNRS/IN2P3, CEA/IRFU, Observatoire de Paris, 119 75205 Paris, France}
 
\author{A.V. Bukevich}
\affiliation{Institute for Nuclear Research of Russian Academy of Sciences, 117312 Moscow, Russia}
 
% ...
\collaboration{500}{(The LHAASO Collaboration)}
 

\correspondingauthor{Y. Luo, C.N. Tong, X.B. Chen, R.Y. Liu, H. Zhou}

\begin{abstract}
Pulsar wind nebula DA~495 (G65.7+1.2) has been extensively observed from radio to TeV $\gamma$-ray bands. We present LHAASO observations of DA~495, revealing an energy-dependent morphology, where an extended source with $r_{39}=0.19^{\circ}\pm0.02^{\circ}$ is detected by WCDA (0.4-15~TeV), and a point-like source with a 95\% upper limit of $r_{39}=0.11^{\circ}$ is observed by KM2A ($>25~\mathrm{TeV}$). The spectrum of the source extends beyond 100~TeV with a break or cutoff at a few tens of TeV. Our X-ray data analysis, based on Chandra and XMM-Newton observations, shows that the X-ray emission of DA~495 extends well to $\sim 6^{\prime}$, significantly larger than the size previously reported. The broadband spectral energy distribution across radio, X-ray and TeV $\gamma$-ray bands is phenomenologically described by a one-zone leptonic model, yielding an average magnetic field of $\sim$ 5 $\mathrm{\mu G}$, while Fermi-LAT spectral analysis indicates a likely presence of a $\gamma$-ray pulsar within the system. A time-dependent model, in which particle transport is convection-dominated in the inner region (within $\sim100^{\prime\prime}$) and diffusion-dominated in the outer region, successfully reproduces the observed radial profiles of X-ray surface brightness and spectral index, and also accounts for the TeV $\gamma$-ray emission detected by LHAASO, suggesting that DA~495 represents an evolved PWN with ongoing particle escape that gives rise to a TeV halo component --- that is, a PWN+halo system.
\end{abstract}


\section{Introduction} \label{sec:Introduction}
Pulsar wind nebula (PWN), a bubble of highly relativistic particles powered by the rotational energy of its pulsar, is known to radiate non-thermal emission across the entire electromagnetic spectrum from $\nu<100~\mathrm{MHz}$ radio waves to $E_{\gamma}>10^{15}~\mathrm{eV}$ photons \citep[PeV emission from the Crab nebula reported by][]{2021Sci...373..425L}. Inside a PWN, the compact, rapidly rotating and highly magnetized neutron star continuously injects relativistic particles including electrons, positrons and possibly nuclei and ions \citep{2021ApJ...922..221L} known as pulsar wind into the ambient medium. The confinement of this pulsar wind in the ambient medium creates a termination shock where these $e^{\pm}$ are randomly oriented and accelerated up to $>10^{15}~\mathrm{eV}$. These $e^{\pm}$ radiate from radio to X-ray in the local magnetic field via synchrotron process, and very-high-energy (VHE; $100~\mathrm{GeV} \leq E_{\gamma} < 100~\mathrm{TeV}$) and ultrahigh-energy (UHE; $100~\mathrm{TeV} \leq E_{\gamma} < 100~\mathrm{PeV}$) $\gamma$-ray via inverse Compton (IC) scattering. As one of the most well-studied kinds of sources in the non-thermal sky, PWNe exhibit diverse morphologies and spectral characteristics due to their rapid temporal evolution that strongly depend on the properties of the central neutron star and the density and structure of the ambient medium into which the relativistic pulsar wind expands. Accordingly, the evolution of PWN can be roughly divided into three stages \citep{2020A&A...636A.113G}: (i) For $t<10~\mathrm{kyr}$, the PWN freely expands within the cold, slow-moving ejecta of the supernova explosion before the reverse shock reaches it. The most well-known PWN at this stage is the Crab nebula \citep{2008ARA&A..46..127H}. (ii) For $t\sim10-100~\mathrm{kyr}$, the PWN enters a more complex evolutionary phase. The reverse shock interacts with the PWN, compressing it and potentially disrupting it. During this stage, the system often exhibits an irregular morphology and these accelerated particles may escape into the supernova remnant (SNR), or even into the surrounding interstellar medium (ISM). A notable example is the Vela-X PWN \citep{2018A&A...617A..78T}. (iii) For $t>~100\mathrm{kyr}$, the neutron star leaves its host SNR, moving supersonically through the ISM and forming a bow-shock PWN. At this stage, these relativistic particles escape and propagate in the ISM, generating a TeV-emitting halo structure (so-called TeV halo or pulsar halo) by up-scattering local background photons including cosmic microwave background (CMB), infrared dust emission and stellar light, etc. The most well-known TeV halos are the Geminga and Monogem halos, first reported by HAWC \citep{2017Sci...358..911A}.

Based on previous observations, DA~495 has been identified as a PWN, but no evidence of pulsations from its associated pulsar has ever been identified. The latest radio analysis of DA~495 was reported by \citet{2008ApJ...687..516K}, revealing a circular diffuse radio emission that decreases gradually to its outer edge with a full extent of $25^{\prime}$ in diameter. The absence of a detectable SNR shell implies that the system either evolved in a very low-density medium or has reached a late evolutionary stage. The magnetic field in the radio nebula was determined to be extremely strong at $1.3~\mathrm{mG}$. In X-ray, previous observations of DA~495 region identified a compact source J1952.2+2925 within a non-thermal extended nebula of $40^{\prime\prime}$ in diameter \citep{2004ApJ...610L.101A,2008ApJ...687..505A}. The compact source J1952.2+2925 was found to have a purely thermal spectrum and identified as an isolated neutron star based on the spectral fit using either a blackbody model or an atmosphere model \citep{2015MNRAS.453.2241K}. These observations confirm the PWN interpretation of DA~495. At TeV energies, DA~495 was firstly detected by HAWC as a point source 2HWC~J1953+294, with a photon index of $-2.78\pm 0.15$ \citep{2017ApJ...843...40A}. This TeV emission was subsequently confirmed by VERITAS as VER~J1952+293. A 2D-Gaussian model was applied to fit VER~J1952+293, revealing that it is extended with $\sigma=0.14^{\circ} \pm 0.02^{\circ}$ and has a photon index of $-2.65 \pm 0.49$ \citep{2018ApJ...866...24A}. The photon indices measured by HAWC and VERITAS are in good agreement, but a significant discrepancy exists between the flux measurements: the flux measured by HAWC at 1~TeV is 7 times higher than that measured by VERITAS \citep{2018ApJ...866...24A}. Due to the significantly smaller size of DA~495 in X-rays compared to its radio/TeV counterpart, a two-zone model was introduced \citep{2019ApJ...878..126C}. Nevertheless, interpreting the multi-wavelength emission of DA~495 remains a challenging task.

In this article, we investigate PWN DA~495 based on LHAASO and multi-wavelength observations. In Section~\ref{sec:LHAASO data analysis}, we report TeV $\gamma$-ray analysis of DA~495 based on LHAASO data \citep[cataloged as 1LHAASO~J1952+2922 in the first LHAASO source catalog,][]{2024ApJS..271...25C}. In Section~\ref{sec:Multi-wavelength observations}, we report an X-ray analysis based on both Chandra and XMM-newton data, together with a high-energy $\gamma$-ray analysis using Fermi-LAT observations. In Section~\ref{sec:Multi-wavelength spectral modeling} and \ref{sec:Discussion}, we discuss the radiation mechanism and explore the properties of DA~495.

\section{LHAASO data analysis} \label{sec:LHAASO data analysis}

\begin{figure}[htbp]
    \centering
    \begin{minipage}{0.48\textwidth}
        \centering
        \includegraphics[width=\textwidth]{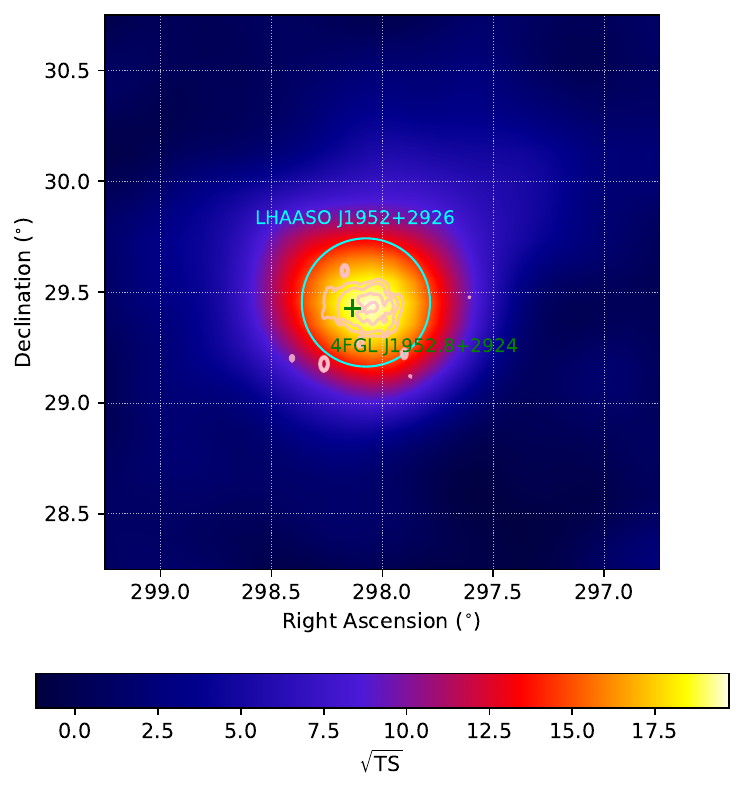}
    \end{minipage}
    \hfill
    \begin{minipage}{0.48\textwidth}
        \centering
        \includegraphics[width=\textwidth]{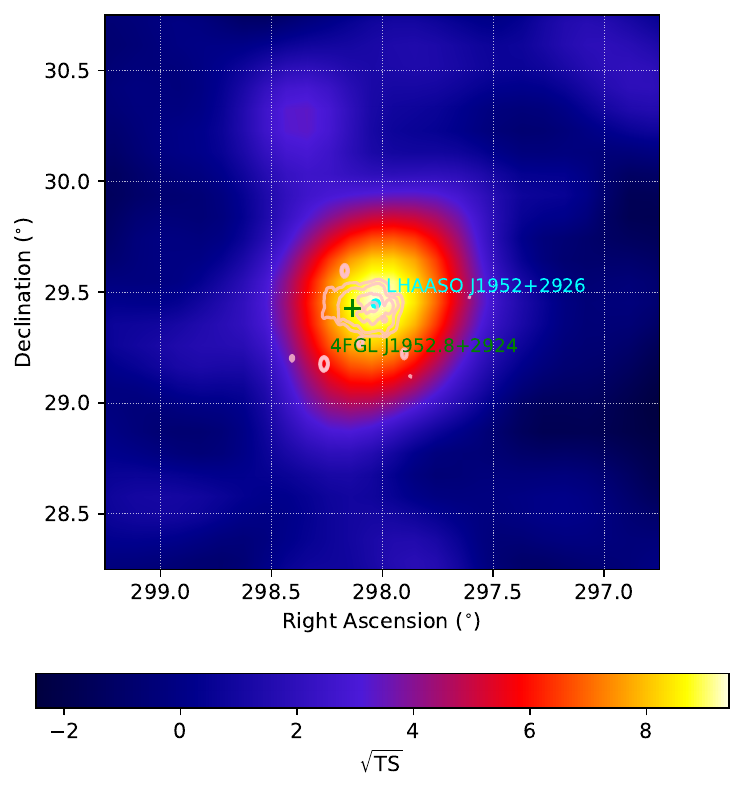}
    \end{minipage}

    \vskip\baselineskip
    \begin{minipage}{0.6\textwidth}
        \centering
        \includegraphics[width=\textwidth]{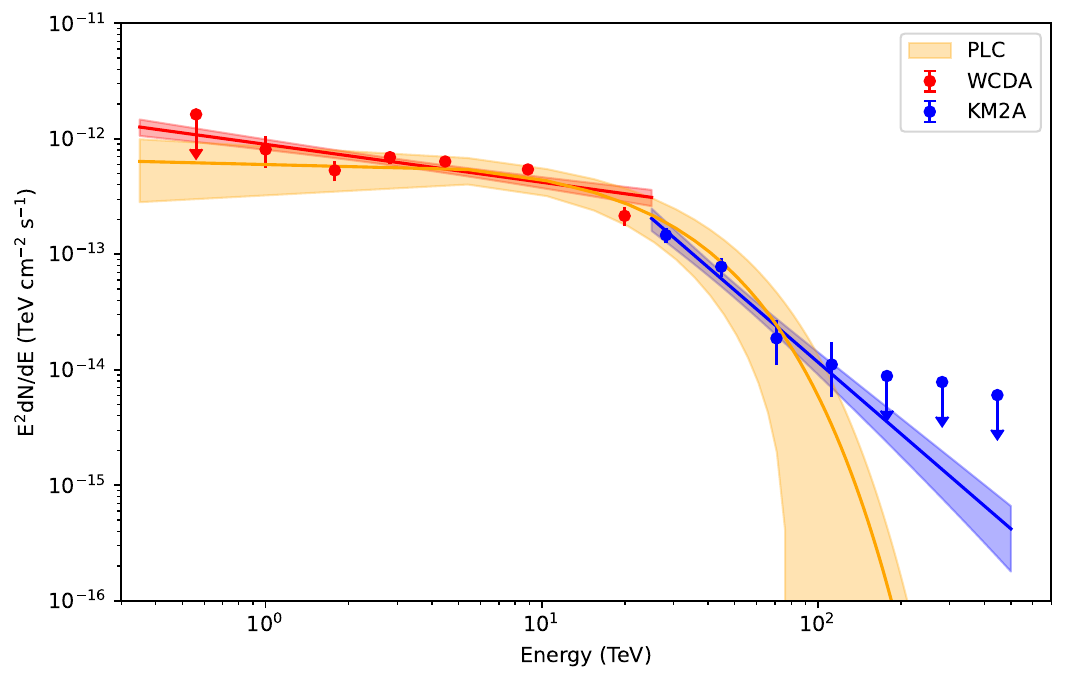}
    \end{minipage}
    \caption{Top panels: significance maps of LHAASO~J1952+2926 region after subtracting nearby sources, obtained with WCDA (0.4-15~TeV; left) and KM2A ($>25~\mathrm{TeV}$; right). LHAASO~J1952+2926 is shown as a cyan circle (68\% containment) in the WCDA map and as a cyan point in the KM2A map. Pink contours indicate $T_b=(8,9,11)~\mathrm{K}$ from the 1420~MHz Canadian Galactic Plane Survey \citep{2003AJ....125.3145T}, and the cross marks 4FGL~J1952.8+2924. Bottom panel: SED of LHAASO J1952+2926. The red and blue points represent WCDA and KM2A measurements, respectively, while the orange line shows the joint PLC fit. The shaded area indicates the $1\sigma$ uncertainty.}
    \label{fig:Significance maps and SED of LHAASO~J1952+2926}
\end{figure}

Both LHAASO-WCDA data and LHAASO-KM2A data were used to investigate TeV $\gamma$-ray emission from DA~495, with a total exposure time of 1137 days (March 2021 to July 2024) and 1569 days (December 2019 to July 2024), respectively. A three-dimensional likelihood framework based on the binned maximum likelihood method was employed to analyze the morphology and spectrum of each source. Given a model with a set of spatial and spectral parameters denoted as $\theta$, likelihood is calculated as
\begin{equation}
    \ln \mathcal{L}(\theta)=\sum_i^{N_{\mathrm{bins}}} \sum_j^{\mathrm{ROI}}\ln P(N_{\mathrm{obs}}^{i,j}|\theta)
\end{equation}
and maximized, where $N_{\mathrm{obs}}^{i,j}$ is the number of observed events in the $j$th pixel of the $i$th energy bin and $P$ is the probability of detecting $N_{\mathrm{obs}}^{i,j}$ events given parameters $\theta$ based on Poisson fluctuation. Likelihood ratio tests were conducted to evaluate how well one model performs better than another. The test statistics (TS), defined as 
\begin{equation}
    \mathrm{TS}=2\ln (\mathcal{L}(\theta_{\mathrm{alt}})/\mathcal{L}(\theta_{\mathrm{null}}))
\end{equation}
is used to compare two nested models, where $\mathcal{L}(\theta_{\mathrm{alt}})$ and $\mathcal{L}(\theta_{\mathrm{null}})$ are the likelihood values obtained from an alternative hypothesis and a null hypothesis respectively. According to Wilks’s theorem \citep{Wilks:1938dza}, TS follows a chi-square distribution with the degree of freedom (dof) being equal to the difference between the number of free parameters of two nested hypotheses. One can convert the TS value into a p-value, and calculate the corresponding pre-trial significance of the alternative hypothesis compared to the null hypothesis, especially $S\simeq\sqrt{\mathrm{TS}}$ when $\mathrm{dof}=1$. 

In this work, we chose a rectangular Region of Interest (ROI) defined by $296^{\circ}<\mathrm{R.A.}<304^{\circ}$ and $28^{\circ}<\mathrm{Decl.}<32^{\circ}$ where R.A. and Decl. are the right ascension and declination respectively. For the nearby diffuse emission component, we employed a spatial template derived from Planck maps of dust optical depth \citep{2014A&A...571A..11P,2016A&A...596A.109P}, as used in the first LHAASO source catalog \citep{2024ApJS..271...25C}. In both WCDA and KM2A data, the TeV counterpart of DA 495 is modeled with a 2D-Gaussian morphology and a power-law spectrum $dN/dE=\Phi_0 (E/{E_{\mathrm{piv}}})^{-\alpha}$, where $E_{\mathrm{piv}}$ is the pivot energy and chosen to be 3~TeV for WCDA data and 50~TeV for KM2A data. During our fitting procedure, PWN DA~495 is significantly detected by WCDA ($\mathrm{TS}=441.53$, corresponding to an energy range of 0.4-15~TeV) and by KM2A ($\mathrm{TS}=120.07$, corresponding to energies $>25~\mathrm{TeV}$), identified as LHAASO~J1952+2926, as shown in the top panels of Figure~\ref{fig:Significance maps and SED of LHAASO~J1952+2926}. In addition to LHAASO~J1952+2926, five other sources also exhibit significant TeV $\gamma$-ray emission, but they are beyond the scope of this work and will not be discussed further in this paper.

The best-fit parameters for LHAASO~J1952+2926 are presented in Table~\ref{tab:Best-fit parameters for LHAASO J1952+2926}. Analysis of WCDA data (0.4-15~TeV) reveals an extended morphology with $r_{39}=0.19^{\circ}\pm0.02^{\circ}$. In contrast, KM2A data ($>25~\mathrm{TeV}$) favor a point-like morphology, and the 95\% statistical upper limit on $r_{39}$ is $0.11^{\circ}$. This energy-dependent size reduction suggests a leptonic origin for the TeV $\gamma$-ray emission, where higher-energy $e^{\pm}$ undergo stronger radiative cooling and consequently propagate shorter distances. The bottom panel of Figure~\ref{fig:Significance maps and SED of LHAASO~J1952+2926} presents the spectral energy distribution (SED) of LHAASO~J1952+2926. A joint fit to WCDA and KM2A data was performed, and the overall spectrum is well described by a power-law with an exponential cutoff (PLC), given by $dN/dE=\Phi_0 (E/10~\mathrm{TeV})^{-\alpha}e^{-E/E_\mathrm{c}}$ with the best-fit parameters listed in Table~\ref{tab:Best-fit parameters for LHAASO J1952+2926}.

\begin{deluxetable*}{lccccccD}[htbp]
\tablenum{1}
\tablecaption{Best-fit parameters for LHAASO~J1952+2926}
\label{tab:Best-fit parameters for LHAASO J1952+2926}
\tablewidth{0pt}
\tablehead{
\colhead{Component} & \colhead{R.A.~($^{\circ}$)} & \colhead{Decl.~($^{\circ}$)} & \colhead{$r_{39}$~($^{\circ}$)} & \colhead{$\Phi_0~(\mathrm{TeV^{-1}~cm^{-2}~s^{-1}})$} & \colhead{$\alpha$} & \colhead{$E_\mathrm{c}~(\mathrm{TeV})$} 
}
\startdata
WCDA & $298.07\pm0.02$ & $29.45\pm0.02$ & $0.19\pm0.02$ &  $(6.93 \pm 0.50) \times 10^{-14}$ & $2.33 \pm 0.07$ & ... \\
KM2A & $298.03\pm0.04$ & $29.45\pm0.04$ & $<0.11$ &  $(1.95 \pm 0.28) \times 10^{-17}$ & $4.07 \pm 0.24$ & ... \\
Joint & $298.05\pm0.02$ & $29.45\pm0.02$ & $0.16\pm0.02$ & $(7.17 \pm 1.67) \times 10^{-15}$ & $1.97\pm0.15$ & $20.59\pm5.02$ 
\enddata
\end{deluxetable*}

\section{Multi-wavelength observations} \label{sec:Multi-wavelength observations}
\subsection{Chandra and XMM-Newton}

\begin{figure}[ht!]
\centering
\raisebox{-0.5\height}{\includegraphics[width=0.45\textwidth]{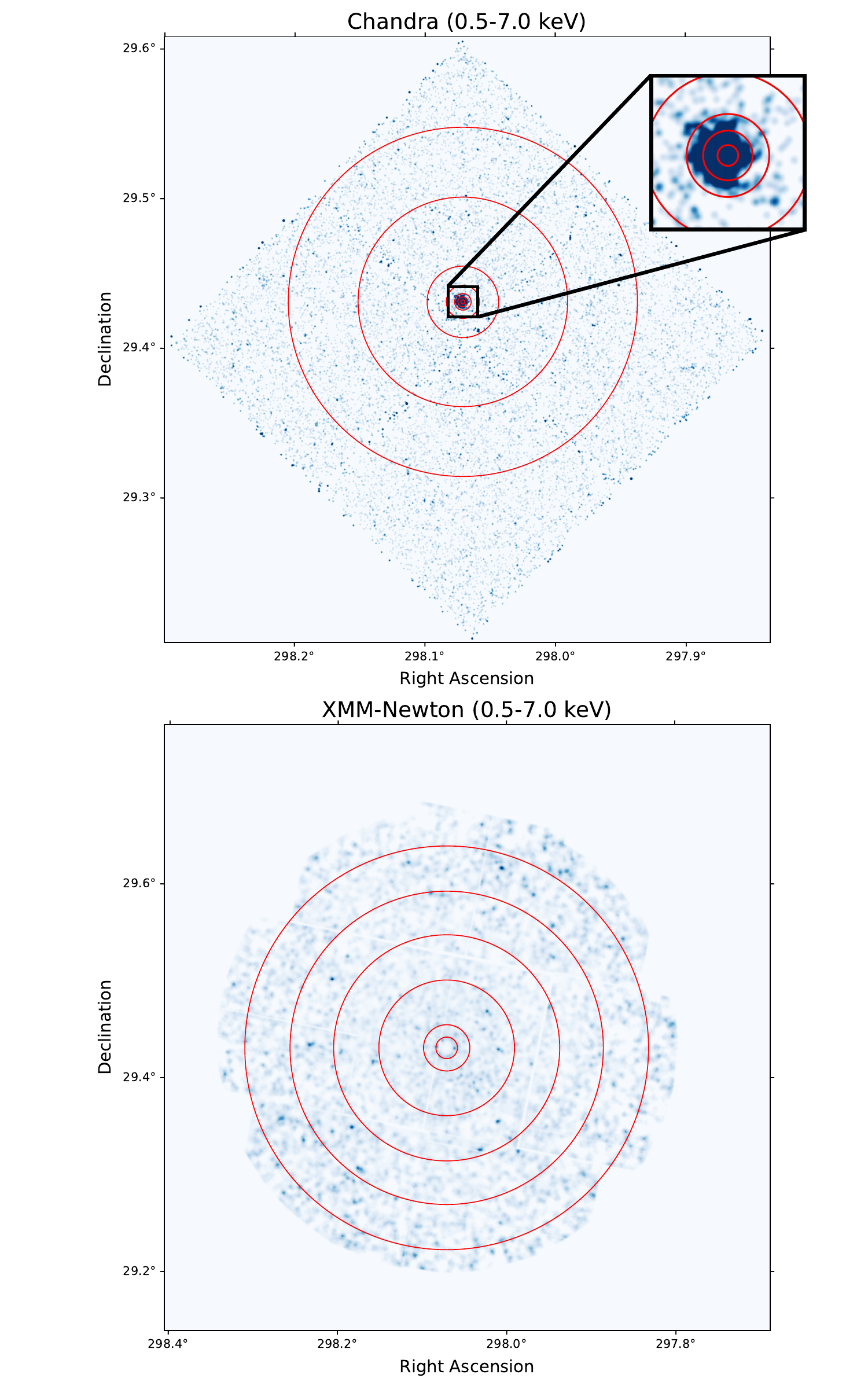}}
\raisebox{-0.5\height}{\includegraphics[width=0.45\textwidth]{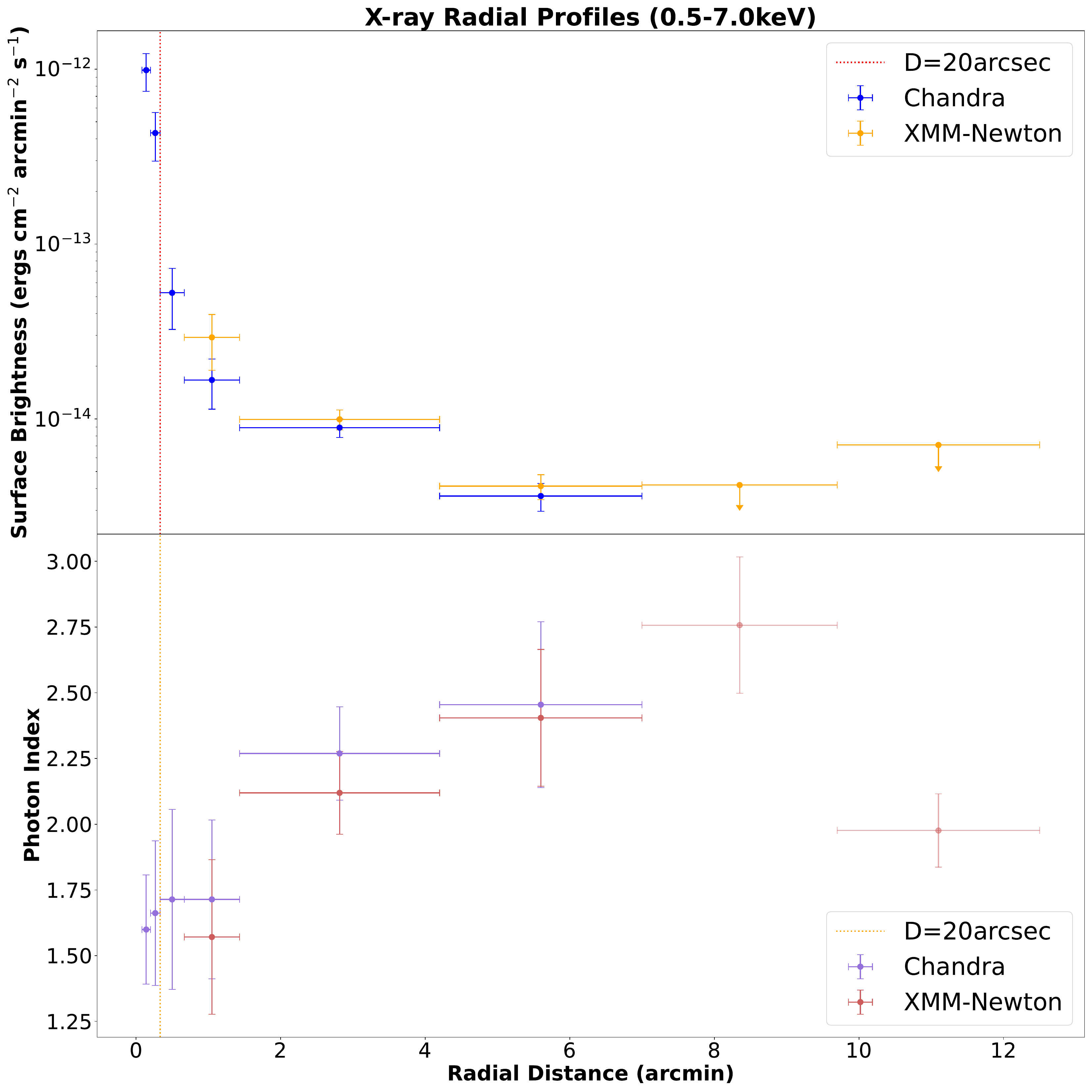}}
\caption{Left panels: Background subtracted and exposure corrected images of X-ray instruments, 0.5-7.0\,keV. Left upper panel: \textit{Chandra} image smoothed with a Gaussian kernel to 3$\sigma$ significance. The point sources were filled with interpolated values from surrounding background regions. The 90$\%$ containment radius at the center of the annuli is $\sim 1.12^{\prime\prime}$.} Left bottom panel: \textit{XMM-Newton} image generated by the \texttt{adapt} tool (\texttt{smoothingcounts=10}) with point sources detected by \texttt{cheese} masked. The 90$\%$ containment radius at the center of the annuli is $\sim 26.94^{\prime\prime}$. The red annular regions in left two panels are used for spatially spectral analysis. Right panels: X-ray radial profiles extracted from annular regions in 0.5-7.0\,keV. The error bars represent the $1\sigma$ uncertainties. Right upper panel: Surface brightness profile, the last two data points overestimate the surface brightness due to the reduced point-source detection capability at the edge of the instrument field of view. Right bottom panel: Index profile. Vertical dashed lines mark the X-ray PWN size with a radius of $20^{\prime\prime}$\citep{2008ApJ...687..505A}.
\label{fig:X-ray image}
\end{figure}

We analyzed the X-ray observation data of DA\,495 performed by \textit{Chandra} and \textit{XMM-Newton}. DA\,495 was observed by \textit{Chandra}/ACIS-I on December 05, 2002 (ObsID 3900). We used \textit{Chandra} Interactive Analysis of Observations (CIAO) version 4.16.0 as well as Calibration Database (CALDB) version 4.11.3 for spatial and spectral analysis. The data were reprocessed using \texttt{chandra\_repro}, and point sources (\(>3\sigma\)) detected by \texttt{wavdetect} in the image produced by \texttt{fluximage} would be masked in later analysis. Background flares (\(>3\sigma\)) were removed using \texttt{deflare} (0.5-7.0\,keV), resulting in cleaned events with an exposure of 24.74\,ks. To subtract the quiescent particle background, we used \textit{Chandra}/ACIS-I stowed background files, which were not exposed to sky. Since the effective area of \textit{Chandra} in the 9.5-12.0\,keV is negligible, essentially all flux in this energy range is contributed by the particle background. Therefore, we scaled the stowed background counts rate in the 9.5-12.0\,keV to match the observation, allowing us to accurately subtract the quiescent particle background. Finally, we used \texttt{dmfilth} to replace the source regions detected by \texttt{wavdetect} with interpolated values from surrounding background regions, producing an image of the real diffuse emission (Figure \ref{fig:X-ray image}). The \textit{XMM-Newton} observation towards DA\,495 was performed on April 21, 2007 (ObsID 0406960101) and the data was processed using the \textit{XMM-Newton} Science Analysis Software (SAS) version 20.0.0 together with the \textit{XMM-Newton} Extended Source Analysis Software (ESAS). The EPIC-PN camera was in Small Window data mode and a CCD of EPIC-MOS1 did not work during the observation, therefore, in the subsequent diffuse emission analysis, we did not use these two cameras and instead only utilized the data of EPIC-MOS2 camera. The raw data was reprocessed by \texttt{emchain}, then the flares were filtered using \texttt{mos-filter} (\texttt{allowsigma=1.5} was applied to effectively remove the soft proton flare background in 0.5-7.0\,keV), resulting in 29.76\,ks cleaned events of EPIC-MOS2. We selected the 0.5–7.0\,keV energy band, the same as used by Chandra, for the image analysis and used \texttt{mos\_spectra} to create the image. The exposure-corrected image with the quiescent particle background (QPB) created by \texttt{mos\_back} subtracted, is shown in Figure~\ref{fig:X-ray image}.

For spectral analysis, we first derived the spatial distribution of the intensity and photon index in the annular regions shown in the left panels of Figure \ref{fig:X-ray image}. We used \texttt{specextract} of CIAO and \texttt{mos\_spectra} of SAS to extract the spectrum in different regions with point sources masked (we also masked point sources detected by \texttt{wavdetect} of \textit{Chandra} in the \textit{XMM-Newton} spectral analysis using circular regions with a radius of $30^{\prime\prime}$), Xspec version 12.13.0c was applied to fit the spectrum. To subtract the local background, we extracted a diffuse sky background spectrum from the ROSAT All-Sky Survey (RASS) data in an annular region centered on the source with inner and outer radii of $0.25^{\circ}$ and $1.25^{\circ}$, which is outside the radio nebula and the energy range of RASS data can well describe the thermal emission from the sky background. The sky background spectrum can be modeled as a combination of a cool ($E\sim$ 0.1\,keV), unabsorbed thermal component originating from the Local Hot Bubble and the heliosphere, a hotter ($E \sim 0.25 - 0.7$\,keV), absorbed thermal emission from the Galactic halo and/or intergalactic medium, and an absorbed power-law component with a photon index of 1.45, accounting for the unresolved cosmic X-ray background from distant extragalactic sources. The source spectrum was modeled with a simple absorbed power-law, with the hydrogen column density ($N_{\mathrm{H}}$) fixed to the value reported by \citet{2019ApJ...878..126C} to better constrain the photon index. We fit the source and background region simultaneously and the radial profiles of intensity and photon index were displayed in the right panels of Figure \ref{fig:X-ray image}. The absorption component of all these models is \textit{tbabs} with abundance \textit{wilm} \citep{2000ApJ...542..914W}. These profiles indicate an extended nebula with emission reaching up to $\sim 6^{\prime}$. This significantly exceeds the previously reported X-ray size of $\sim 20^{\prime\prime}$ in radius \citep{2008ApJ...687..505A}, suggesting a much larger X-ray nebula\footnote{\citet{2008ApJ...687..505A} directly subtracted the background from the edge of field of view, where the detection efficiency is lower (vignetting), thereby underestimating the background counts. \citet{2015MNRAS.453.2241K} used background regions inside the radio nebula and the gamma-ray emission area, which are almost filled with the Chandra and XMM-Newton field of view, thus likely overestimating the background counts. \citet{2019ApJ...878..126C} employed background modeling rather than background subtraction, which avoids vignetting effect. However, they considered only an extragalactic background component and neglected the Galactic background contribution. In our analysis, in addition to the extragalactic background component, we modeled the Galactic background with ROSAT All-Sky Survey data, selecting regions outside the radio/gamma-ray emission area.}. Consequently, we obtained an integrated flux from $5^{\prime\prime}$ to $12.5^{\prime}$ (shown in Figure~\ref{fig:Multi-wavelength SED of PWN DA~495}) and the two-zone model previously proposed \citep{2019ApJ...878..126C} to account for the small ($\sim 20^{\prime\prime}$ in radius) X-ray nebula is no longer necessary.

\subsection{Fermi-LAT}
According to the latest Fermi-LAT point source catalog 4FGL-DR4 \citep{2023arXiv230712546B}, a high-energy (HE; $100~\mathrm{MeV} \leq E_{\gamma} < 100~\mathrm{GeV}$) $\gamma$-ray source 4FGL~J1952.8+2924 is likely to be correlated with DA~495 with an offset of $\sim8^{\prime}$. Therefore, we utilized approximately 17 years of Fermi-LAT Pass 8 data collected between 2008 August~4 and 2025 October~28, covering energies from 100~MeV to 1~TeV, to analyze the $\gamma$-ray emission around 4FGL~J1952.8+2924 using Fermipy package \citep{2017ICRC...35..824W}. The photon events were constrained to those with event class \verb|"P8R3_SOURCE"| (evclass=128) and event type FRONT+BACK (evtype=3). As the Fermi-LAT team recommended, we adopted the standard data quality selection criteria \verb|"DATA_QUAL>0 && LAT_CONFIG==1"|, and events with zenith angle $\geq 90^{\circ}$ were also excluded to avoid contamination from the earth limb. The ROI was chosen to be $15^{\circ} \times 15^{\circ}$ centered at the position of 4FGL~J1952.8+2924. The source models were constructed using 4FGL-DR4 catalog. The Galactic and extragalactic diffuse emission were modeled using the files \verb|gll_iem_v07.fits| and \verb|iso_P8R3_SOURCE_V3_v1.txt| respectively. The instrument response functions (IRF) \verb|"P8R3_SOURCE_V3"| were used for performing binned maximum likelihood analysis. 4FGL~J1952.8+2924 is localized at $\mathrm{R.A.}=298.13^{\circ}\pm0.02^{\circ}$ and $\mathrm{Decl.}=29.43^{\circ}\pm0.01^{\circ}$, as shown in Figure~\ref{fig:Significance maps and SED of LHAASO~J1952+2926}. The source appears point-like, with no significant extended emission detected. The obtained $\gamma$-ray fluxes are presented in Figure~\ref{fig:Multi-wavelength SED of PWN DA~495}, represented by gray squares.

\section{Multi-wavelength spectral modeling} \label{sec:Multi-wavelength spectral modeling}
Based on these multi-wavelength observations of DA~495, it is evident that DA~495 is an energetic PWN emitting non-thermal radiation from radio to $\gamma$-rays above 100~TeV. We attempt to interpret these emission with a one-zone leptonic model, where the observed radio to X-ray emission arises from synchrotron process and the TeV $\gamma$-ray emission are produced via IC scattering mechanism, both driven by the same population of $e^{\pm}$. No hadronic component is expected, as there is no indication suggesting its necessity.

\begin{figure}[tbp]
    \centering
    \includegraphics[width=0.8\textwidth]{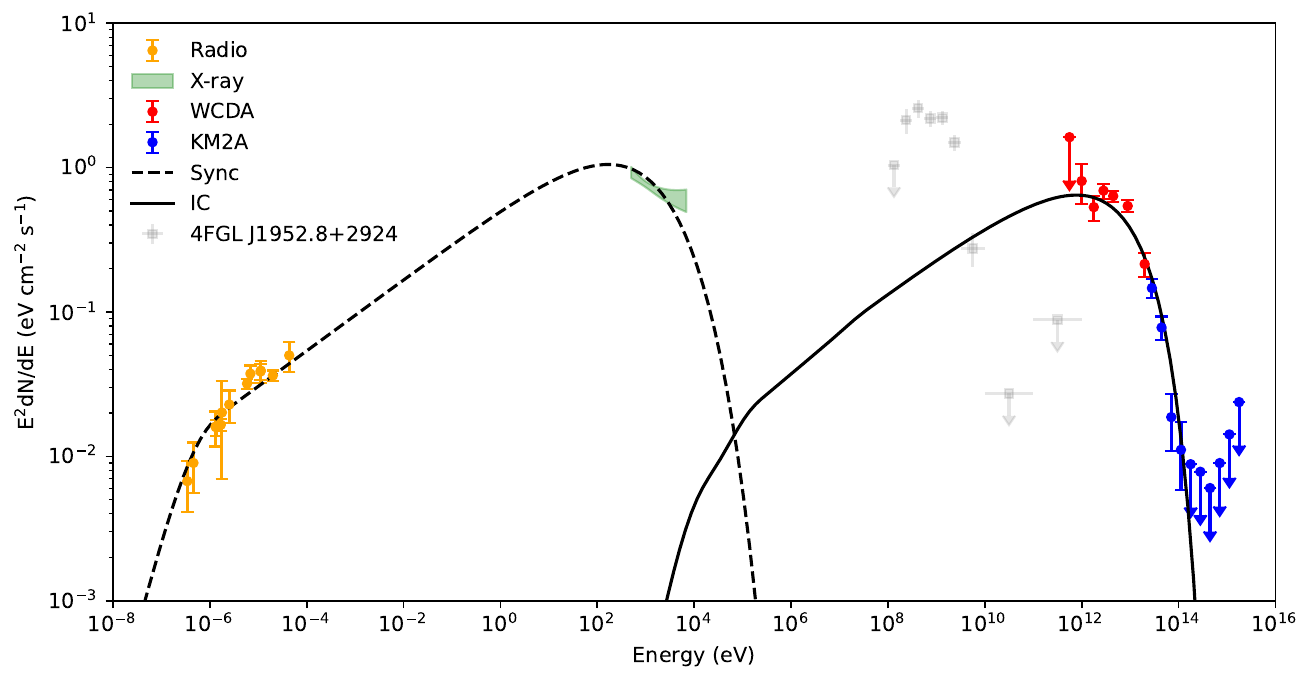} 
    \caption{Multi-wavelength SED of PWN DA~495 including radio, X-ray, and TeV $\gamma$-ray fluxes. Here we adopted the radio fluxes reported by \citet{2008ApJ...687..516K}. The $\gamma$-ray fluxes of 4FGL~J1952.8+2924 is shown as gray squares. The radio, X-ray, and TeV $\gamma$-ray emission from the nebula were modeled through synchrotron process and IC scattering mechanism, represented by the dashed and solid lines respectively.} 
    \label{fig:Multi-wavelength SED of PWN DA~495} 
\end{figure}

With this one-zone leptonic model, a phenomenological fit was performed (see Figure~\ref{fig:Multi-wavelength SED of PWN DA~495}) using the Python package NAIMA \citep{2015ICRC...34..922Z}. The distance to DA~495 in the fit was fixed at 1~kpc as reported by \citet{2008ApJ...687..516K}.
For the target photon field, we assumed CMB ($T=2.73~\mathrm{K}$, $U_{\mathrm{ph}}=4.21\times10^{-13}~\mathrm{erg~cm^{-3}}$), far-infrared radiation ($T=30~\mathrm{K}$, $U_{\mathrm{ph}}=9.27\times10^{-13}~\mathrm{erg~cm^{-3}}$), near-infrared radiation ($T=500~\mathrm{K}$, $U_{\mathrm{ph}}=2.86\times10^{-13}~\mathrm{erg~cm^{-3}}$) and visible light ($T=5000~\mathrm{K}$, $U_{\mathrm{ph}}=8.11\times10^{-13}~\mathrm{erg~cm^{-3}}$) derived from the model of \citet{2017MNRAS.470.2539P}) to be up-scattered by the population of relativistic $e^{\pm}$, which was described by a power-law with an exponential cutoff function, defined as
\begin{equation}
    dN_{\mathrm{e}}/dE_{\mathrm{e}}=\Phi_{\mathrm{e}}(E_{\mathrm{e}}/1~\mathrm{TeV})^{-\alpha_{\mathrm{e}}}e^{-E_{\mathrm{e}}/E_{\mathrm{e,c}}}
\end{equation}
. The best-fit values are $B=5.46\pm0.12~\mathrm{\mu G}$, $\Phi_{\mathrm{e}}=(1.15\pm0.06)\times10^{33}~\mathrm{eV^{-1}}$, $\alpha_{\mathrm{e}}=2.51\pm0.01$ and a cutoff energy of $E_{\mathrm{e,c}}=55.62\pm3.53~\mathrm{TeV}$.

\section{Discussion} \label{sec:Discussion}
\subsection{a $\gamma$-ray pulsar?} \label{sec:a gamma-ray pulsar}
As mentioned in Section~\ref{sec:Introduction}, pulsations from the putative pulsar J1952.2+2925 within DA 495 have yet to be detected. An isolated neutron star detected in X-ray provides a direct evidence supporting the pulsar scenario, however, our understanding of this putative pulsar remains highly limited, as we lack knowledge of its period, surface magnetic field, age and spin-down power, etc. Fermi-LAT detected a point source 4FGL J1952.8+2924 consistent with the position of the neutron star. Its spectrum exhibits a significant cutoff at GeV energies (as shown in Figure 3), resembling the conventional spectrum of pulsed emission from $\gamma$-ray pulsars. This suggests that the $\gamma$-ray source 4FGL J1952.8+2924 may associate with the putative pulsar J1952.2+2925. Its absence from the Fermi-LAT third catalog of $\gamma$-ray pulsars \citep[3PC,][]{2023ApJ...958..191S} is probably attributed to insufficient photon statistics for pulsation detection. If future observations with longer exposure and improved timing analysis confirm a definitive pulsation detection, they will provide critical constraints on the age and energetics of this system.

\subsection{spin-down power}
Based on Hillas criterion \citep{1984ARA&A..22..425H}, the maximum energy of particles accelerated inside a PWN--regardless of the specific acceleration mechanism--can be related to the pulsar's spin-down power and estimated as $E_{max}\approx2\eta_e\eta_B^{1/2}\dot{E}_{36}^{1/2}~\mathrm{PeV}$ \citep{2022ApJ...930L...2D}, where $\eta_e$ is the acceleration efficiency \citep[$\eta_e\leq1$ under the ideal magnetohydrodynamics (MHD) condition,][]{2002PhRvD..66b3005A}, $\eta_B$ is the fraction of spin-down energy converted into magnetic energy at the termination shock ($\eta_B\leq1$), and $\dot{E}_{36}$ is the spin-down power in units of $10^{36}~\mathrm{erg/s}$. Given that the observed $\gamma$-ray spectrum extends up to $E_{\gamma}\approx112~\mathrm{TeV}$, the corresponding parent $e^{\pm}$ population should reach energies of about $E_e\approx399~\mathrm{TeV}$, according to the empirical relation between parent $e^{\pm}$ energy and IC-upscattered photon energy, $E_e\simeq2.15E_{\gamma,15}^{0.77}~\mathrm{PeV}$  \citep[with $E_{\gamma,15}$ in units of $10^{15}~\mathrm{eV}$,][]{2021Sci...373..425L}. At the very least, $e^{\pm}$ must be accelerated to 112~TeV to produce the highest-energy photons observed. Consequently, even assuming the unrealistic maximum efficiency ($\eta_e=1$, $\eta_B=1$), the spin-down power of the central pulsar in DA 495 must satisfy $\dot{E}\gtrsim4\times10^{34}~\mathrm{erg/s}$ ($3\times10^{33}~\mathrm{erg/s}$) to account for $e^{\pm}$ acceleration with $E_{e}\approx399~\mathrm{TeV}$ ($112~\mathrm{TeV}$).

\subsection{Time-dependent modeling of DA~495}
While the one-zone leptonic model provides a useful first-order description of the broadband spectral energy distribution (SED) of DA~495, it fails to capture the spatial features and spectral softening seen in the X-ray regime. To account for these aspects, we further develop a time-dependent model based on the spherically symmetric transport equation that governs the evolution of relativistic $e^\pm$ within the PWN.

We follow the framework of \citet{2024ApJ...976..172C}. In a spherically symmetric system, the transport of particles with number density $n = n(r, \gamma, t)$ within the nebula can be written as the Fokker–Planck equation \citep{1965P&SS...13....9P}: 
\begin{equation} \label{eq:dndt}
\frac{\partial n}{\partial t} = 
D \frac{\partial^2n}{\partial r^2} + 
\left[\frac1{r^2}\frac\partial{\partial r}(r^2D)-V\right]\frac{\partial n}{\partial r} - 
\frac1{r^2}\frac\partial{\partial r}[r^2V]n + 
\frac\partial{\partial\gamma}[\dot{\gamma}n]+Q_{\rm inj}.
\end{equation}
In the above equation, $\dot{\gamma}$ is the summation of particle energy losses, including the adiabatic expansion loss, synchrotron radiation energy loss and IC scattering energy loss, $D$ denotes the diffusion coefficient, $V$ is the bulk velocity of electrons i.e. the convection velocity, and the function $Q_{\rm inj}$ is the distribution of particles injected from the termination shock. For the energy spectrum of electrons freshly injected into the nebula we assume the following power-law shape $n \propto \gamma^{-\alpha}$. The total energy carried by relativistic electrons is assumed to be a fraction $\eta_e$ of the  spin-down power. The maximum Lorentz factor of electrons is estimated as $\gamma_{\max}=\frac{4\epsilon e}{m_ec^2}\sqrt{c\eta_B\eta_e\dot{E}}$.
The radial profile of the magnetic field inside the nebula can be given by $B(r,t) = B_0(t)\left( {r}/{R_{\rm ts}(t)} \right) ^{-\beta_B}  + 3\mu G$, with the total magnetic energy assumed to be a fraction $\eta_B$ of the  spin-down power.

A single-zone or purely diffusive model cannot simultaneously reproduce the steep decline of the X-ray surface brightness and the broad TeV morphology of DA~495.
To better capture the observed morphology of DA~495, we hance adopt a hybrid transport model in which the electron propagation is governed by convection in the inner region and by diffusion in the outer region. 
This approach is motivated by the expectation of strong magnetic turbulence and bulk motion near the termination shock, which suppresses diffusion and favors advection, while the outer region becomes more diffusion-dominated due to reduced pressure and more ordered fields.

We define the transition radius $r_b$ between these two regimes at approximately 100 arcseconds, based on the break observed in the X-ray surface brightness profile. The transport coefficients are set as:
\begin{equation}\label{eq:vr}
v(r) =\left\{ 
\begin{array}{ll}
v_0\left(\dfrac{r}{r_b}\right)^{-\beta} & \text{for } r < r_b \quad \text{(convection zone)}\\
0 & \text{for } r \geq r_b
\end{array}\right.
\end{equation}
and
\begin{equation}\label{eq:Diff}  
D(E) =\left\{ 
\begin{array}{ll}
0 & \text{for } r < r_b \\
D_0 \left(\dfrac{E}{100\ \mathrm{TeV}}\right)^\delta & \text{for } r \geq r_b \quad \text{(diffusion zone)}
\end{array} \right.
\end{equation}
for a typical Kolmogorov turbulence diffusion $\delta=1/3$ \citep{1941DoSSR..30..301K}.

This two-zone transport model enables a more realistic description of the multi-wavelength morphology. Electrons injected by the pulsar are advected outward within the inner convective region, losing energy primarily via synchrotron radiation, resulting in the compact X-ray nebula. Beyond the transition radius, electrons enter the diffusion-dominated regime where they spread more slowly and upscatter ambient photons to TeV energies, generating the observed extended $\gamma$-ray halo \citep{2012ApJ...752...83T, 2011ApJ...742...62V}.

\begin{figure}
    \centering
    \includegraphics[width=0.45\textwidth]{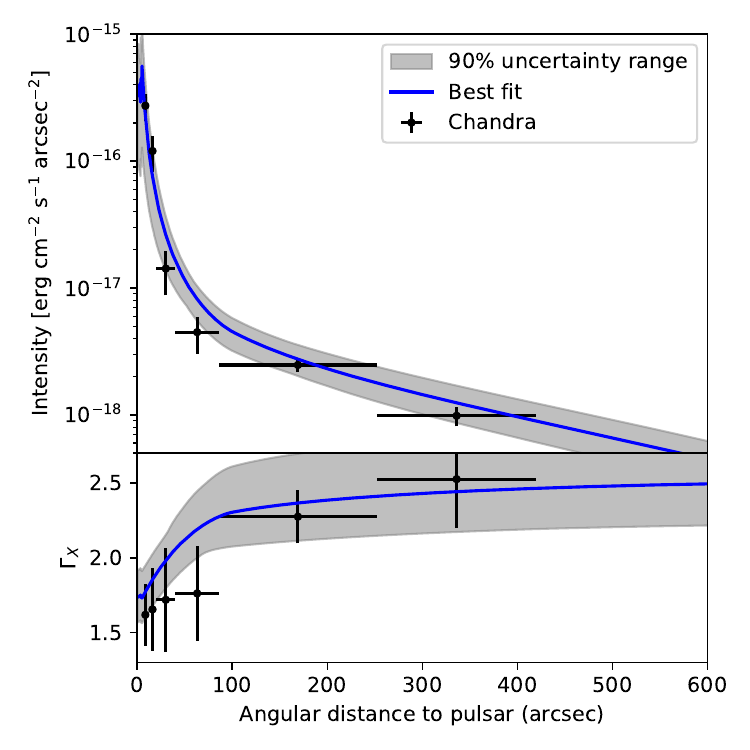}
    \includegraphics[width=0.45\textwidth]{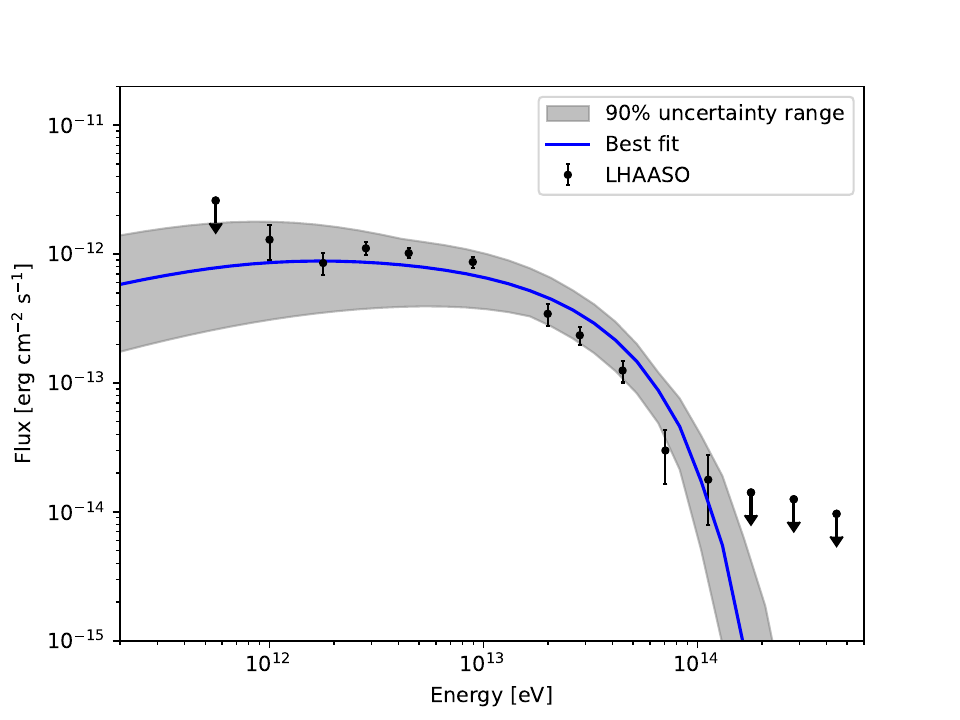}
    \caption{Left panel: Radial profile of the X-ray surface brightness and photon index as a function of angular distance from the pulsar. The blue curves represent the best-fit model predictions, incorporating a transition from convection- to diffusion-dominated transport at $\sim 100^{\prime\prime}$. Right panel: Modeled $\gamma$-ray spectrum (blue line) compared to the LHAASO observations of DA~495. The model reproduces both the flux level and spectral cutoff in the multi-TeV range, consistent with inverse Compton emission from relic electrons. The error bars in both panels indicate the 90\% uncertainties of the best-fit results derived from the Markov Chain Monte Carlo (MCMC) sampling of the model parameter space.}
    \label{fig:model_X-gamma}
\end{figure}

We perform a Markov Chain Monte Carlo (MCMC) fitting procedure for constraining the key transport and injection parameters. In the modeling, several source quantities are fixed, including the pulsar spin-down power ($\dot{E}=10^{36}\mathrm{erg\,s^{-1}}$), the initial convective velocity ($v_0=0.3c$), the radial index of the velocity profile ($\beta=0.8$), and the initial diffusion coefficient ($D_0=10^{28}\ \mathrm{cm^{2}\ s^{-1}}$). These fixed parameters are chosen based on typical values inferred for evolved PWNe and previous modeling efforts (e.g., \citealt{2024ApJ...976..172C,Peng2022ApJ...926....7P}). The adopted convective velocity ($v_0=0.3c$) corresponds to the post-shock flow speed expected from relativistic MHD simulations, while the diffusion coefficient ($D_0=10^{28},\mathrm{cm^2,s^{-1}}$) represents a canonical value for the interstellar medium around TeV PWNe. Reasonable variations in these fixed parameters have little impact on the overall results and conclusions. Therefore, fixing them allows the MCMC to focus on the key physical parameters that govern the radiative evolution.
The free parameters determined by the MCMC fitting are the magnetic fraction ($\log\eta_B=-2.68^{+0.12}_{-0.12}$), electron fraction ($\log\eta_e=-0.53^{+0.27}_{-0.21}$), the acceleration efficiency parameter ($\epsilon=0.89^{+0.12}_{-0.11}$), the injection spectral index ($\alpha=2.24^{+0.09}_{-0.08}$), and the magnetic-field radial index ($\beta_B=0.65^{+0.05}_{-0.05}$). 
As shown in the left panel of Figure~\ref{fig:model_X-gamma}, the radial profile of the X-ray surface brightness and spectral index are well reproduced by our model, which assumes a convection-dominated zone within $\sim100^{\prime\prime}$ and a diffusion-dominated outer region. The photon index increases with distance from the pulsar, consistent with the scenario of synchrotron cooling.
In addition, our model successfully explains the TeV $\gamma$-ray emission observed by LHAASO (the right panel of Figure~\ref{fig:model_X-gamma}) through inverse Compton scattering of the same electron population. The predicted spectral shape and flux level are in good agreement with the observed data, further supporting the evolutionary scenario of DA~495 as a relic PWN with ongoing particle escape and cooling.

Furthermore, we examined whether the predicted spatial extension is consistent with the TeV observation. From the modeled two-dimensional flux intensity distribution, we derived an intrinsic radius enclosing 39\% of the total flux, obtaining $r_{39} \simeq 0.15^{\circ}$, which is close to the value measured by WCDA. Physically, TeV emission cannot be confined solely to the nebular center: particles are efficiently transported outward by convection and diffusion on timescales shorter than their radiative cooling, while the inverse Compton target photon field remains nearly homogeneous. As a result, the TeV radiation is naturally extended.

\section{Conclusion} \label{sec:conclusion}
In this work, we present a comprehensive multi-wavelength analysis of PWN DA~495, combining TeV $\gamma$-ray observations from LHAASO, X-ray data from Chandra and XMM-Newton, along with Fermi-LAT $\gamma$-ray measurements. The LHAASO data reveal an energy-dependent TeV morphology and a spectrum extending beyond 100~TeV. The X-ray analysis indicates the X-ray emission extends to $\sim6^{\prime}$, significantly larger than previously reported. Fermi-LAT spectral analysis suggests the presence of an as-yet-unidentified $\gamma$-ray pulsar within the system. While a one-zone leptonic model phenomenologically explains the broadband spectrum across radio, X-ray and TeV $\gamma$-ray bands, a two-zone particle transport model better reproduces the morphologies, indicating DA~495 is an evolved PWN with a developing TeV-halo component.

\begin{acknowledgments}
We would like to thank all staff members who work at the LHAASO site above 4400 meter above the sea level year round to maintain the detector and keep the water recycling system, electricity power supply and other components of the experiment operating smoothly. We are grateful to Chengdu Management Committee of Tianfu New Area for the constant financial support for research with LHAASO data. We appreciate the computing and data service support provided by the National High Energy Physics Data Center for the data analysis in this paper. This research work is supported by the following grants: The National Natural Science Foundation of China No.12393853, No.12393852, No.12393851, No.12393854, NSFC No.12205314, No.12105301, No.12305120, No.12261160362, No.12105294, No.U1931201, No.12375107, NSFC No.12173039, the Department of Science and Technology of Sichuan Province, China No.24NSFSC2319, Project for Young Scientists in Basic Research of Chinese Academy of Sciences No.YSBR-061, and in Thailand by the National Science and Technology Development Agency (NSTDA) and the National Research Council of Thailand (NRCT) under the High-Potential Research Team Grant Program (N42A650868).
\end{acknowledgments}

\bibliography{export-bibtex}
\bibliographystyle{aasjournal}

\end{document}